\documentclass[aip, amsmath, amssymb, reprint]{revtex4-1}

\usepackage{graphicx}% Include figure files
\usepackage{dcolumn}% Align table columns on decimal point
\usepackage{bm}% bold math
\usepackage{docmute}
\usepackage{color}

\usepackage[utf8]{inputenc}
\usepackage[T1]{fontenc}
\usepackage{mathptmx}
\usepackage{etoolbox}

\makeatletter
\def\@email#1#2{%
 \endgroup
 \patchcmd{\titleblock@produce}
  {\frontmatter@RRAPformat}
  {\frontmatter@RRAPformat{\produce@RRAP{*#1\href{mailto:#2}{#2}}}\frontmatter@RRAPformat}
  {}{}
}%
\makeatother

\begin{document}

\preprint{AIP/123-QED}

\title[Sample title]{Polarization analysis of terahertz emission from Bi-2212 cross-whisker intrinsic Josephson junctions and its refractive index}
% Force line breaks with \\

\author{Y. Saito}
\email{SAITO.Yoshito@nims.go.jp}
\affiliation{Graduate School of Pure and Applied Sciences, University of Tsukuba, 1-1-1 Tennodai, Tsukuba, Ibaraki 305-8577, Japan}
\affiliation{International Center for Materials Nanoarchitectonics (MANA), National Institute for Materials Science, Tsukuba, Ibaraki 305-0047, Japan}

\author{I. Kakeya}
\affiliation{Department of Electronic Science and Engineering, Kyoto University, Nishikyo, Kyoto 615-8510, Japan}

\author{Y. Takano}
\affiliation{Graduate School of Pure and Applied Sciences, University of Tsukuba, 1-1-1 Tennodai, Tsukuba, Ibaraki 305-8577, Japan}
\affiliation{International Center for Materials Nanoarchitectonics (MANA), National Institute for Materials Science, Tsukuba, Ibaraki 305-0047, Japan}

\date{\today}

\begin{abstract}

	Polarization analyses of the terahertz (THz) emission from Bi$_{2}$Sr$_{2}$CaCu$_{2}$O$_{8+\delta}$ whisker crystals used for superconducting THz emitters were conducted.
	The THz emission mode was estimated by a simple polarization measurement, and a simulation study was conducted to examine the validity of the polarization analysis. 
	The refractive index of whisker crystals revealed through the polarization analyses was greater than that of bulk single crystals and agreed well with our previous THz emission report.   
	The simulation study suggested the complex plasma excitation mode of the THz emission, and an interpretation of the refractive index obtained in this study is provided.

\end{abstract}

\maketitle

%********************************************************************************************************************************************************
%***********************************************************Introduction***********************************************************************************
%********************************************************************************************************************************************************

	A superconducting terahertz (THz) emitter taking advantage of intrinsic Josephson junctions (IJJs) in high-temperature superconductor Bi$_{2}$Sr$_{2}$CaCu$_{2}$O$_{8+\delta}$ (Bi-2212) 
	has attracted interest because of its capability of continuous THz wave generation at a frequency of approximately 1 THz in the $\mu$W range \cite{Kleiner1994, Ozyuzer2007}.
	(see Refs. \cite{Kakeya2016, Kashiwagi2017, Kleiner2019} for recent reviews).
	Polarization measurement and control of the radiated THz wave from Bi-2212 devices have been of recent interest.  
	The mesa-structure shape determines the polarization because the surface superconducting current oscillates in the THz frequency. 
	Thus, the device's structure initiates an essential property of a THz source.
	Previously, we reported on the THz emission from a device made of Bi-2212 whisker crystals instead of bulk single crystals \cite{Saito2021}.
	Whisker crystals are small, needle-like single crystals suitable for micro-electronics applications \cite{Cattaneo2021, Kubo2010}.
	In our previous report, a detailed analysis of the THz emission was not possible due to the unknown THz emission mode and refractive index $\sqrt{\varepsilon_{r}}$. 
	Assuming that the refractive index of whisker crystals is possibly different from that of bulk single crystals due to its non-stoichiometric composition ($\sim$ Bi$_{3}$Sr$_{2}$Ca$_{2}$Cu$_{3}$O$_{x}$),
	it should be obtained through emission mode identification as explained below \cite{Nagao2001, Kishida2001}.
	
	The THz emission is governed by the cavity resonance formula $f_{mp} = (c_{0}/2\sqrt{\varepsilon_{r}})\sqrt{(m/W)^2+(p/L)^2}$,
	where $c_{0}$ is the speed of light in a vacuum; $W$ and $L$ are the shorter and longer sides of a rectangular sample, respectively;
	and $m$ and $p$ are the numbers of nodes of a standing wave along the $W$ and $L$ directions, respectively \cite{Kadowaki2008, Klemm2011}.
	The conventional value of $\sqrt{\varepsilon_{r}}$ = 4.2 for Bi-2212 bulk single crystals was deduced from the linear scaling relation of the emission frequency and 1/$W$,
	implying that the THz emission takes place in the transverse mode with zero wavenumbers along the $L$ direction.
	Thus, while the THz emission mode has been conventionally determined from sample dimensions and the emission frequency,
	it can be more directly estimated from the polarization direction of the emitted electromagnetic waves.

	In this study, we conducted a polarization measurement for several THz emitting samples made of Bi-2212 whisker crystals with varied dimensions.
	The refractive index of whisker crystals is obtained from the THz emission mode estimated through polarization analysis and found to be greater than that of Bi-2212 bulk single crystals.
	In addition, the validity of the polarization analysis is investigated by a simulation study.
	The simulation results suggest that the polarization features are not explainable by a single-mode plasma excitation at the resonance point,
	which is consistent with recent experimental and theoretical studies \cite{Benseman2019, Kobayashi2022(2)}. 

%*********************************************************************************************************************************************
%***************************************************************Experimental*******************************************************************
%*********************************************************************************************************************************************

	\begin{figure}
		\includegraphics[width=75mm]{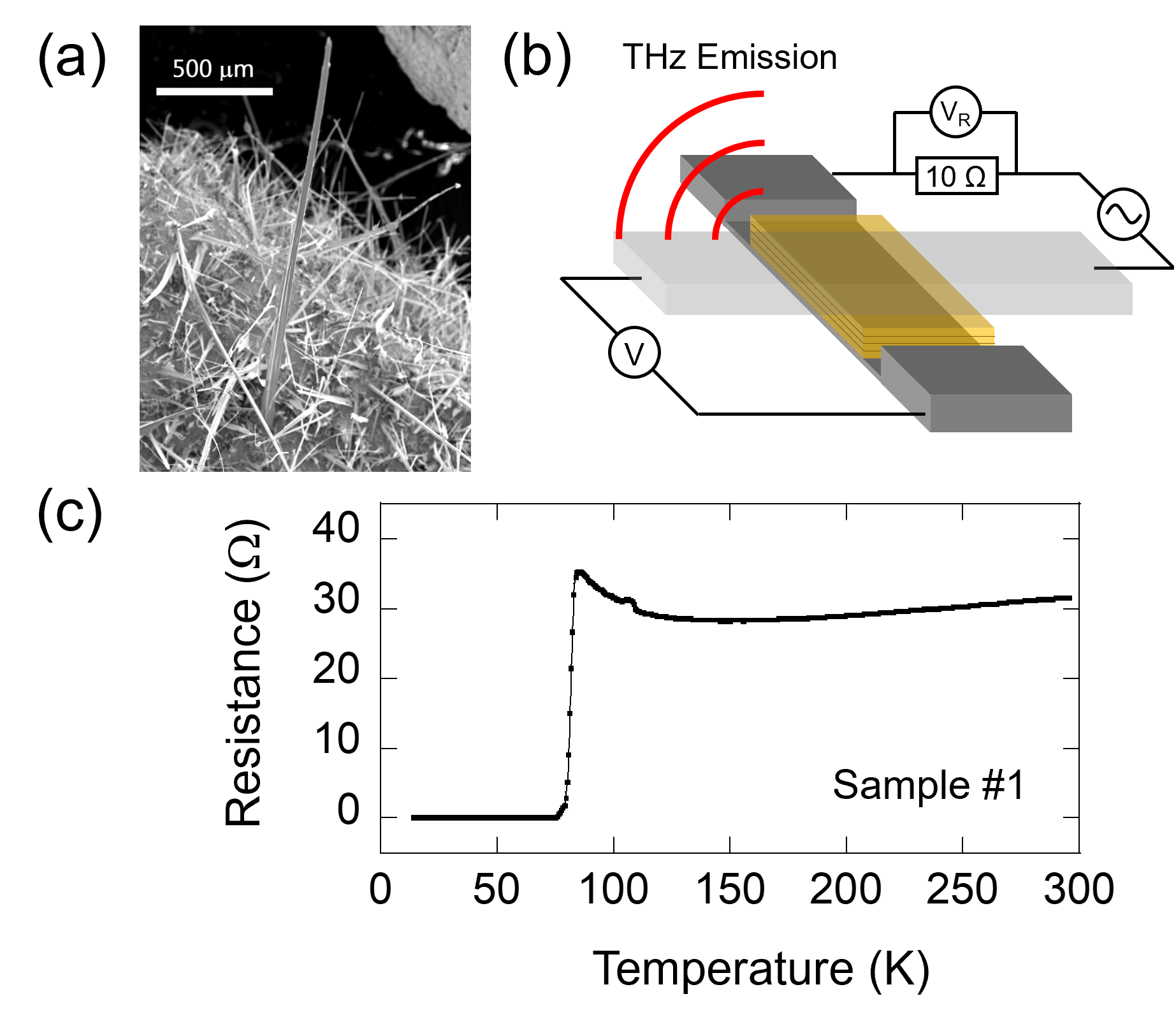}
		\caption{(a) Scanning electron microscope image of typical Bi-2212 whiskers grown using Te-doping method. (b) Schematic of CW-IJJs sample with associated electrical diagram. 
		(c) Temperature dependence of resistance for sample 1.}
	\end{figure}

	Figure 1(a) shows a scanning electron microscope image of Bi-2212 whisker crystals grown using the Te-doping method \cite{Nagao2001}.
	In this method, whisker crystals are grown from precursor pellets containing Te and excess Ca elements by oxygen annealing in a tubular furnace \cite{Saito2022}.
	The typical composition of the whiskers was Bi$_{2.4}$Sr$_{1.6}$Ca$_{1.4}$Cu$_{2}$O$_{x}$, as analyzed by energy-dispersive X-ray spectroscopy.
	The stacks of IJJs were fabricated in a lower whisker of "cross-whisker junction" samples (see Fig. 1(b)), hereinafter referred to as CW-IJJs \cite{Takano2001, Takano2002}.
	A detailed fabrication procedure can be found in the literature \cite{Saito2021}.
	Table I summarizes the characteristics of the samples.

	The temperature dependence of the resistance ($R$-$T$) for sample 1 is shown in Fig. 1(c);
	it exhibits a slight upturn with a nearly unity ratio of $R$($T_{\rm c}$)/$R$(300 K) and a decrease of the resistance to $\sim$ 150 K,
	suggesting that the whisker is in a slightly over-doped state. 
	
	Figure 2 shows the experimental setup for the optical measurements.
	The THz emission was collimated by a high-density polyethylene lens and then transported into a liquid He-cooled InSb hot electron bolometer through an optical chopper and mirror. 
	The spectrum of the THz wave was obtained using a Fourier transform infrared (FT-IR) spectrometer (JASCO Inc., FARIS-1) with a resolution limit of 0.25 cm$^{-1}$.
	Polarization measurements were conducted as follows. 
	First, the sample was current-biased at a certain emission point.
	A wire-grid polarizer (WGP) placed in the optical path was rotated from 0$^{\circ}$ to 360$^{\circ}$ in steps of 10$^{\circ}$ or 15$^{\circ}$.
	The extinction ratio of the WGP reached 25 dB at 1.0 THz. 
	The bolometer signal was recorded to accumulate data at each WGP angle. 
	After the polarization measurement was completed, FT-IR spectroscopy was performed.
	The bolometer signal in the non-emission state was also recorded to determine the background signal level, which is essential for determining polarization.

	\begin{table}
	\caption{Dimensions of CW-IJJs samples, $T_{\rm c}^{\rm Zero}$, bath temperature $T_{\rm b}$ and frequency $f$ where polarization was measured, and estimated cavity resonance mode.}
		\begin{ruledtabular}
			\begin{tabular}{cccc}
				Parameters & Sample 1 & Sample 2 & Sample 3 \\
				\hline
				Width ($\mu$m) & 40 & 48 & 50 \\
				Length ($\mu$m) & 80 & 61 & 117 \\
				Height ($\mu$m) & 0.9 & 0.8 & 0.8 \\
				$T_{\rm c}^{\rm Zero}$(K) & 73 & 72 & 61 \\
				$T_{\rm b}$(K) & 40 & 40 & 55 \\
				$f$ (GHz) & 737 & 488 & 264 \\
				Mode & TM$_{10}$ & TM$_{01}$ & TM$_{01}$  
			\end{tabular}
		\end{ruledtabular}
	\end{table}

	\begin{figure}
		\includegraphics[width=75mm]{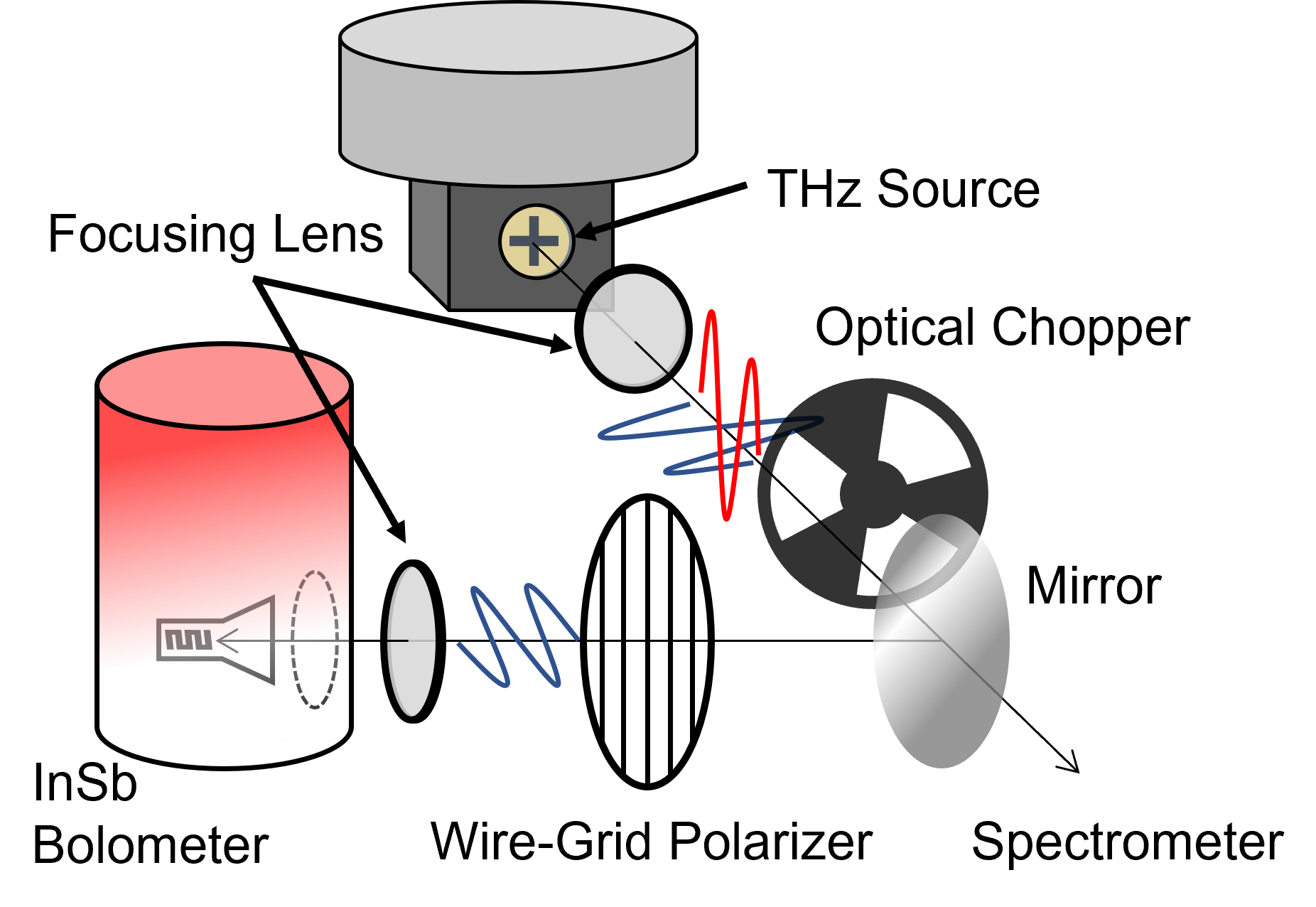}
		\caption{A schematic of experimental setup for polarization and Fourier transform infrared spectroscopy measurement.}
	\end{figure}

%************************************************************************************************************************************************
%***************************************************************Results&Discussion*****************************************************************
%************************************************************************************************************************************************

	\begin{figure}
		\includegraphics[width=75mm]{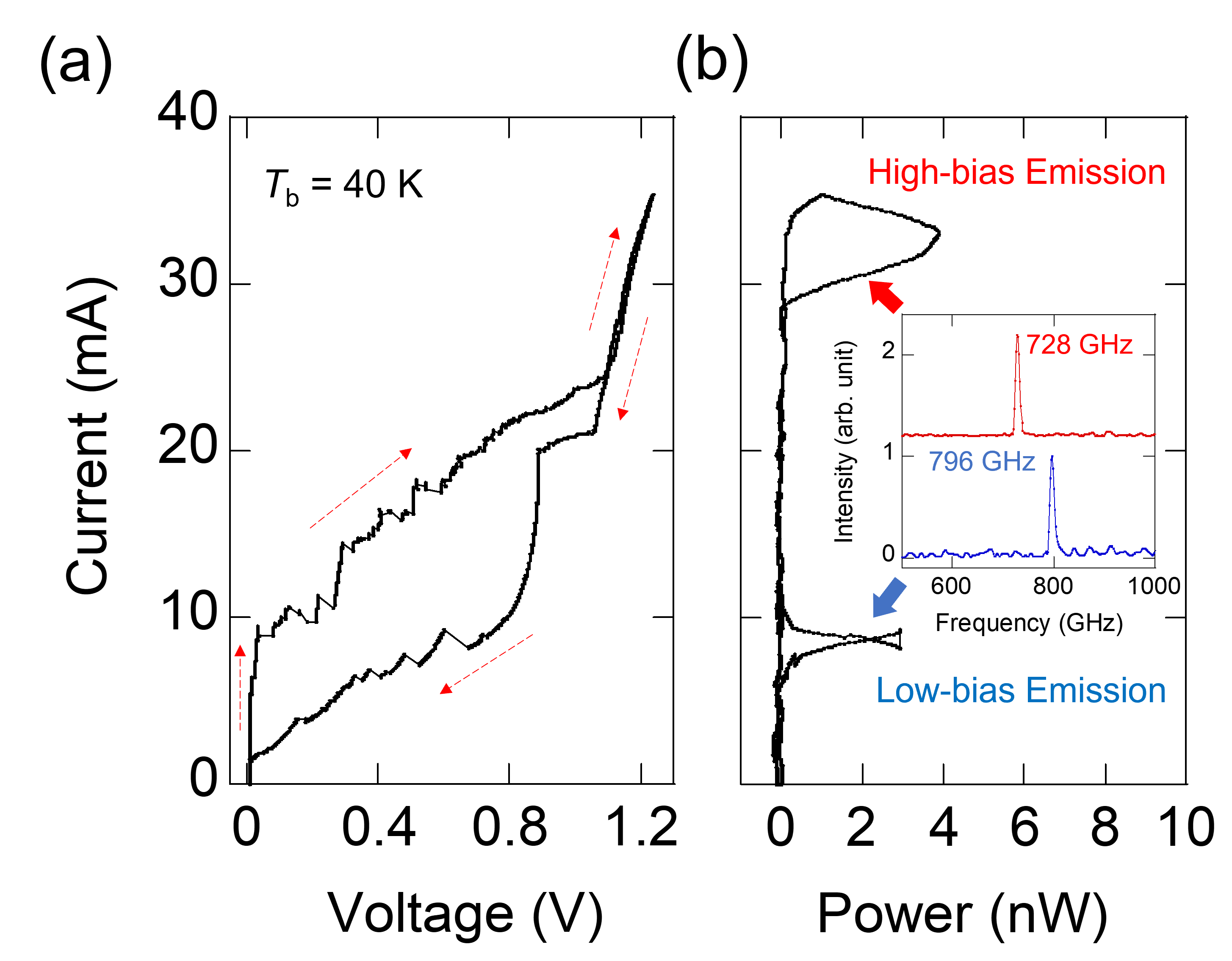}
		\caption{(a) Current-voltage characteristics (IVCs) of sample 1 at bath temperature $T_{\rm b}$ = 40 K. Dashed red arrows indicate current-bias sweep. 
		(b) Bolometer signal corresponding to IVCs. Inset shows Fourier transform infrared spectrum for high-bias (red) and low-bias (blue) regimes.}
	\end{figure}

	Figure 3(a) shows the current-voltage characteristics (IVCs) of sample 1 at a bath temperature of $T_{\rm b}$ = 40 K. 
	The IVCs showed hysteresis, which is typically observed in the underdamped IJJ array. 
	The IJJ system started to become resistive at a current bias of $I \approx$ 10 mA, 
	and the critical current gradually increased to $I \approx$ 25 mA before the entire IJJ system became resistive.
	The back-bending feature in the IVCs was less distinguishable than that of previous studies;
	it has been reported that this feature originates from the self-heating effect of an IJJ mesa \cite{Yurgens2011, Demirhan2015}.
	For our samples, a good cooling efficiency was expected for the following reasons: 
	1) the sample volume was smaller than that of typical IJJ mesas ($\sim$ 400 $\times$ 80 $\times$ 1 $\mu$m) and 
	2) whereas typical IJJ mesas are cooled through the Bi-2212 base crystal with a thickness $>$ 10 $\mu$m, our CW-IJJ samples were thermally connected directly to a MgO substrate. 
	The thermal conductivity of Bi-2212 is one thousandth that of MgO \cite{Slack1962, Fujishiro1994}.
	In addition, the superconducting top electrode of CW-IJJs might generate less Joule heating than metallic electrodes,
	contributing to the efficient cooling performance.

	\begin{figure*}
		\centering
		\includegraphics[width=135mm]{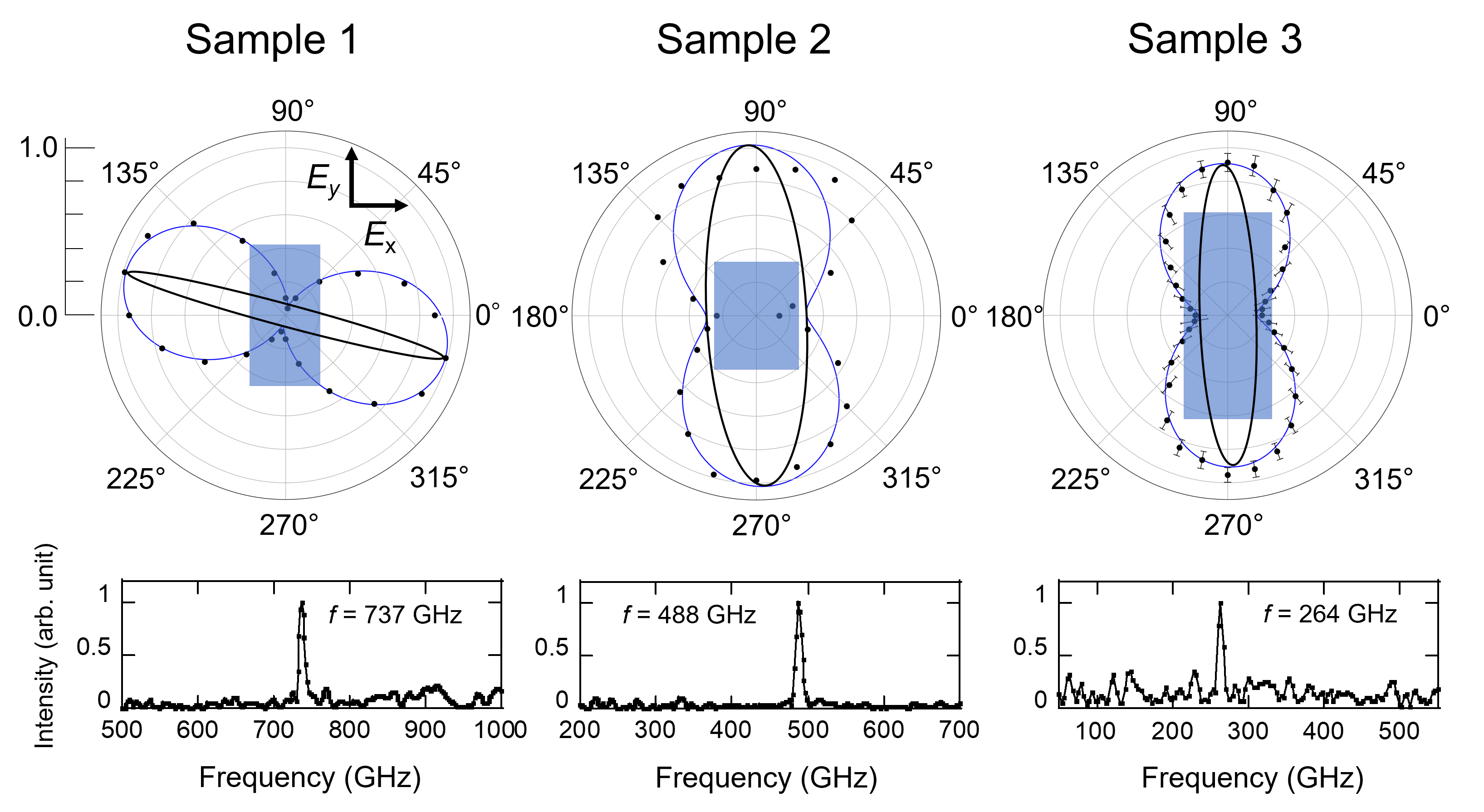}
		\caption{Polarization plot and Fourier transform infrared spectrum for samples 1 - 3. Black dots represent wire-grid polarizer transmission intensity in terms of rotation angle $\theta$.}
	\end{figure*}

	Figure 3(b) displays the lock-in signal of the bolometer signal corresponding to Fig. 3(a), 
	indicating that the THz emission occurs in both high-bias ($I$ = 24-35 mA) and low-bias regimes ($I <$ 10 mA) \cite{Kakeya2016}.
	The inset of Fig. 3(b) shows the normalized FT-IR spectra obtained at the maximum intensity of each emission regime. 
	Radiation in the range of 700-740 GHz for the high-bias regime and 790-890 GHz for the low-bias regime was confirmed by FT-IR spectroscopy.

	The polarization properties of the THz emission are now analyzed to identify the oscillation mode.
	The refractive index of the Bi-2212 whiskers is then discussed.
	Based on previous studies, the THz emission originates from the high-frequency current on the sample surface driven by the standing wave of synchronized transverse Josephson plasma, 
	which is governed by cavity resonance excitation.
	To discuss the cavity resonance mode, we obtained the emission frequency and polarization profile using a WGP.
	The rotation angle $\theta$ was defined as the angle between the metal wire of the WGP and the reference direction.
	Here, the reference direction ($\theta$ = 0) was parallel to the longer side (length) of the sample, which was defined as the $y$-axis.
	The $x$-axis was defined in the same manner as the shorter side (width) of the sample.
	Because a polarization ellipse represents the angular dependence of the detected power ($I(\theta) \propto |{\bm E}|^{2}$),
	it can be decomposed into diagonal components of electric fields: 
	$E_{x} = E_{0x}\exp[i(kz-\omega t+\delta_{x})]$ and $E_{y} = E_{0y}\exp[i(kz-\omega t+\delta_{y})]$ 
	where $E_{0x}$ and $E_{0y}$ are the amplitudes of the electric fields, and $\delta_{x,y}$ are arbitrary phases.

	The cavity resonance mode can be roughly estimated from the electric field amplitude ratio $|E_{x}|/|E_{y}| = E_{0x}/E_{0y}$.
	For example, the predominance of $|E_{x}|$ over $|E_{y}|$ suggests the internal cavity mode excitation along the sample width, 
	namely TM$_{10}$ mode as the lowest excitation mode.
	Electric fields are not directly measurable but can be inferred by analyzing the polarization ellipse using the Jones formulation.
	The angle-dependent power $I(\theta)$ for fully polarized light passing through a linear polarizer is expressed as \cite{Collet2005}

	\begin{align}
		I(\theta) = E_{0x}^{2}\cos^{2}{\theta} + E_{0y}^{2}\sin^{2}{\theta} + 2E_{0x}E_{0y}\cos{\theta}\sin{\theta}\cos{\delta}.
	\end{align} 

	Thus, the value of $|E_{x}|/|E_{y}| = E_{0x}/E_{0y}$ can be obtained from the fitting parameters of $E_{0x}$, $E_{0y}$, and $\delta (= \delta_{x} - \delta_{y})$. 
	It should be noted that the degree of polarization of the raw emission remains uncertain because the Jones formulation can only be applied to fully polarized light.
	However, the magnitude relation between $E_{0x}$ and $E_{0y}$ was not affected by the incoherent components.

	Figure 4 shows polar plots of the WGP transmission intensity and corresponding FT-IR spectra for samples 1 - 3, which were simultaneously obtained at a specific fixed bias point. 
	It is noteworthy that this experiment was enabled by the stable emission in the high-bias regime.
	In the polar plots, the black dots represent the normalized bolometer signal amplitude, and the polarization ellipses are shown as black lines. 
	The error bar for sample 3 corresponds to the 2$\sigma$ error.
	The solid blue lines represent the least-squares fitting of Eq. (1), where the pale blue rectangles represent the aspect ratio of the sample.
	The emission modes were estimated from the fitting results $E_{0x}/E_{0y}$ and by assuming that these were the fundamental modes of TM$_{01}$ or TM$_{10}$ modes.
	We obtained $E_{0x}$/$E_{0y}$ = 2.7 (TM$_{10}$ mode) for sample 1, $E_{0x}$/$E_{0y}$ = 0.54 (TM$_{01}$ mode) for 2,  and $E_{0x}$/$E_{0y}$ = 0.44 (TM$_{01}$ mode) for 3.
	In this case, the lowest excitation (TM$_{01}$) was not observed for sample 1.
	We attributed this to a mismatch in the ac-Josephson relation.

	\begin{figure}
		\includegraphics[width=75mm]{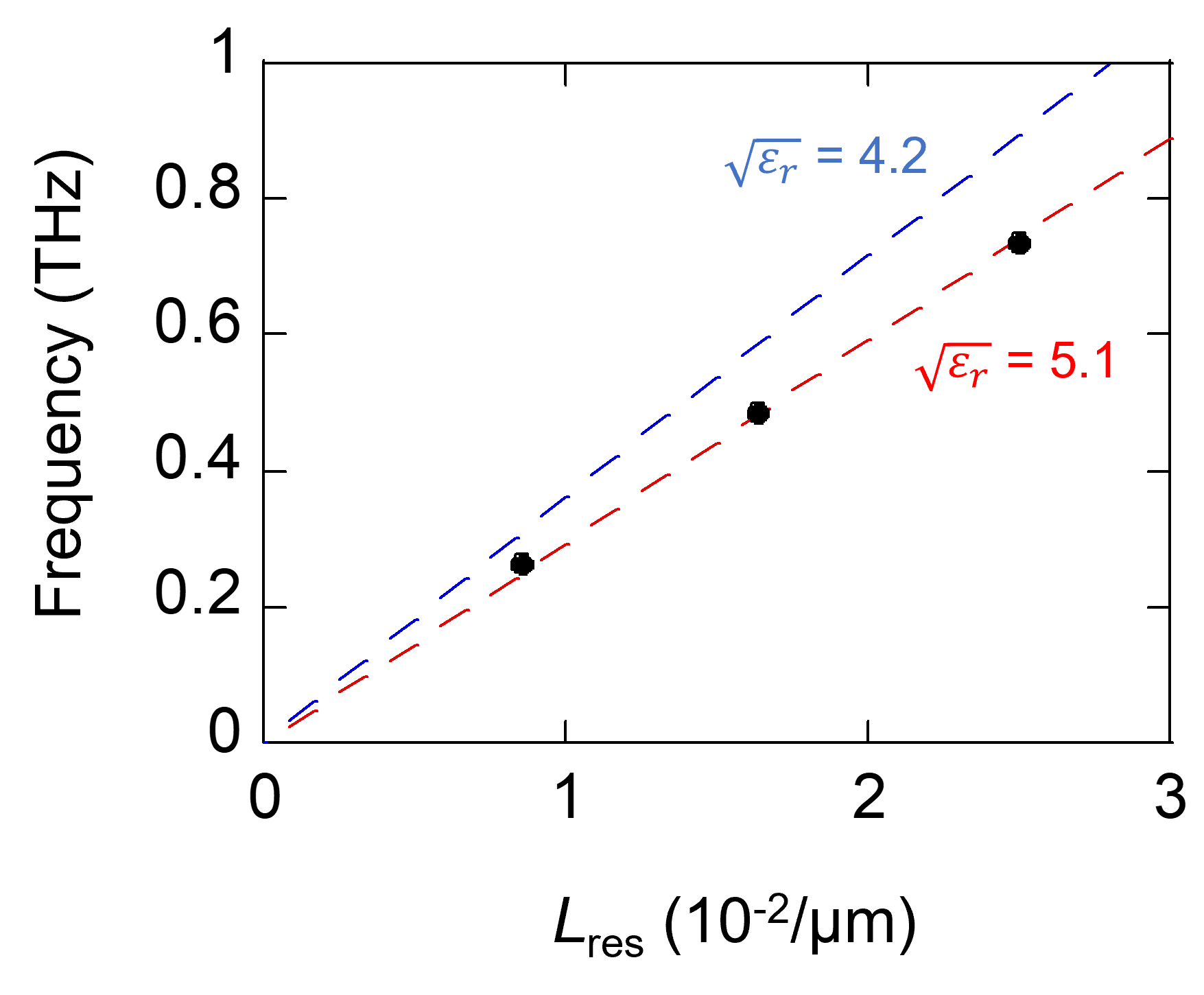}
		\caption{Relationship of emission frequency and inverse of cavity resonant length ($L_{\rm res}$) deduced from polarization measurement.}
	\end{figure} 

	Figure 5 shows a frequency plot of the inverse of the cavity resonant length $L_{\rm res}$ identified from the mode identification.
	In the half-wavelength (TM$_{01}$ or TM$_{10}$) mode, the cavity resonance formula is reduced to $f = c_{0}/2\sqrt{\varepsilon_{r}}L_{\rm res}$.
	Thus, the emission frequency is inversely proportional to the resonant length $L_{\rm res}$, and the proportionality factor reflects the refraction index $\sqrt{\varepsilon_{r}}$.
	This plot is equivalent to the asymptotic line of the dispersion relation for transverse Josephson plasma, $\omega(k)/\omega_{p}=\sqrt{1+\lambda_{c}^{2}k^{2}}$,
	where $\omega_{p}$ is the Josephson plasma frequency and $\lambda_{c}$ is the penetration length along the $c$-axis \cite{Revcolevschi1996}.
	When wavenumber $k$ is sufficiently large, the dispersion relation is approximated by the linear relationship $\omega(k) = \omega_{p}\lambda_{c}k$.
	By considering the formula $\omega_{p} = c_{0}/\sqrt{\varepsilon_{r}}\lambda_{c}$, the linear relationship $\omega(k) = (c_{0}/\sqrt{\varepsilon_{r}})k$ can be obtained,
	which coincides with the cavity resonance formula.
	The dielectric constant $\varepsilon_{r}$ discussed here is the high-frequency dielectric constant of the insulating layers \cite{Kakeya1998}.

	The experimental data for samples 1 - 3 are plotted in Fig. 5 as black dots, and the red dotted line represents linear fitting.
	From the fitting results, the refractive index of Bi-2212 whisker crystals was found to be $\sqrt{\varepsilon_{r}}$ = 5.1.
	This value has good agreement with our previous THz emission report by assuming that it was in the TM$_{02}$ mode,
	where the sample dimensions were 85 $\times$ 30 $\mu$m in-plane, and the maximum emission was obtained at $f \sim$ 680 GHz \cite{Saito2021}.

	The obtained refractive index of whisker crystals ($\sqrt{\varepsilon_{r}}$ = 5.1) was greater than that of bulk single crystals ($\sqrt{\varepsilon_{r}}$ = 4.2, Blue dotted line in Fig. 5)
	\cite{Kadowaki2008, Kashiwagi2011}.
	This change in the refractive index could be attributed to the difference in the material composition:
	Bi$_{2.4}$Sr$_{1.6}$Ca$_{1.4}$Cu$_{2}$O$_{x}$ for our whisker crystals and Bi$_{2.1}$Sr$_{1.7}$Ca$_{1.0}$Cu$_{2}$O$_{x}$ for typical bulk single crystals \cite{Mochiku1994}.
	Conversely, optical response studies have reported a refractive index of  $\sqrt{\varepsilon_{r}} \approx$ 3.5 for Bi-2212 single crystals.
	This value was obtained from both $c$-axis far-infrared reflectivity and microwave resonance measurement \cite{Motohashi2000, Tajima1993, Gaifullin2000}.
	The difference between the refractive index derived from the THz emission study and optical response is possibly due to the difference 
	in nature between spontaneous excitation and the passive response of Josephson plasma in Bi-2212 system.
	
	We now discuss the simulation study conducted to examine the validity of the mode identification, which was roughly estimated from $E_{0x}/E_{0y}$.
	Since we have assumed the two lowest orthogonal modes of TM$_{01}$ and TM$_{10}$, linear polarization in the parallel or vertical direction should be expected.
	However, the experimental results (Fig. 4) show the polarization ellipses with an offset of the main axis (sample 1) and a small axial ratio (samples 2-3).
	Here, we show that these features can be reproduced by setting an appropriate configuration in 3D electromagnetic simulation software (CST STUDIO SUITE).
	The inset of Fig. 6 shows the simulation of sample 1 modeled as a perfect electric conductor (PEC) antenna patch atop the dielectric layer ($\varepsilon_{r}$ = 26), with a thickness of 0.9 $\mu$m. 
	The in-plane size of the patch was 40 $\times$ 80 $\mu$m.
	We used a PEC in the modeling and simulation because the experimental value of the surface impedance in Bi-2212 at the THz region was unknown,
	and we assumed that PEC modeling was not a decisive factor in the polarization simulation.
	The feeding line was also modeled as a PEC patch placed at the corner of the antenna patch, with an area of 10 $\times$ 5 $\mu$m$^{2}$. 
	The position of the feeding line was chosen to have good resonance quality (-9 dB at 730 GHz, full width at half maximum $<$ 2 GHz) at the proximity of the emission frequency ($f$ = 737 GHz). 
	Figure 6 shows simulated traces of the normalized electric fields at nearby frequencies.
	The linear polarization along the $x$-axis at $f$ = 732 GHz corresponds to the antenna resonance of this model,
	and the experimental polarization ellipse of sample 1 (shown in Fig. 4) is reproduced between $f$ = 735 and 739 GHz.
	Subsequently, the polarization direction deviates from the $x$-axis, and a small axial ratio appears as the frequency increases.
	This is consistent with the offset of the polarization measurement point from the intensity maximum of sample 1.
	Recently, Kobayashi et al. proposed an LCR (inductor/capacitor/resistor) resonant circuit model for a stack of IJJs to describe the experimental results
	by demonstrating the mismatch between the resonant frequency and frequency of the THz emission intensity maximum \cite{Kobayashi2022(2)}.

	One interpretation of this elliptical polarization is that the feed position was diagonally placed on a rectangular patch.
	According to antenna theory, the phase difference between orthogonal excitation modes, which creates a circular polarization, is generated by placing a feed point diagonally on a rectangular patch.
	This can be interpreted as the transverse Josephson plasma being excited dominantly in the TM$_{01}$ or TM$_{10}$ mode in our IJJ system, along with another orthogonal mode with a phase difference.
	Previously, Elarabi et al. demonstrated circular polarization from truncated square mesas and notched circular mesas
	by taking advantage of the perturbation effect between two orthogonal modes with a phase difference of $\pi$/2 rad \cite{Elarabi2017, Elarabi2018}.
	Their work proved that Josephson plasma was simultaneously excited in two orthogonal cavity modes with phase differences under appropriate conditions.
	Although the perturbation effect is hardly expected from rectangular IJJ mesas with large aspect ratios ($L/W >$ 4),
	Benseman et al. explained that low-temperature scanning laser microscopy patterns can be approximated as the sum of two cavity modes with a phase difference, forming a composite mode excitation
	\cite{Wang2010, Benseman2019}.
	In addition, Stokes parameter analysis showed small axial ratios and a non-negligible phase difference $\delta_{xy}$ from rectangular IJJ mesas \cite{Tsujimoto2020}.
	In conjunction with previous studies, our polarization measurement results agree with the composite-mode excitation model for Josephson plasma in a cavity.
	Since whisker crystals have an extremely elongated shape, typically $>$ 1 mm $\times$ 30 $\mu$m in the $ab$-plane,
	this composite-mode excitation would be excluded by choosing the width of CW-IJJs too short to be a resonant length. 

	The discussed nature of the THz emission might originate from the spontaneous excitation of Josephson plasma,  
	unlike a conventional patch antenna driven by an external alternating current source.
	This is consistent with the earlier discussion explaining the difference between the two types of refractive index using the different nature of spontaneous and passive Josephson plasma excitation.
	Thus, the refractive index and polarization features discussed in this report are the results of the unique nature of the spontaneous Josephson plasma excitation in the Bi-2212 IJJ system.
	However, further polarization studies are necessary for a more detailed discussion.

	\begin{figure}
		\includegraphics[width=75mm]{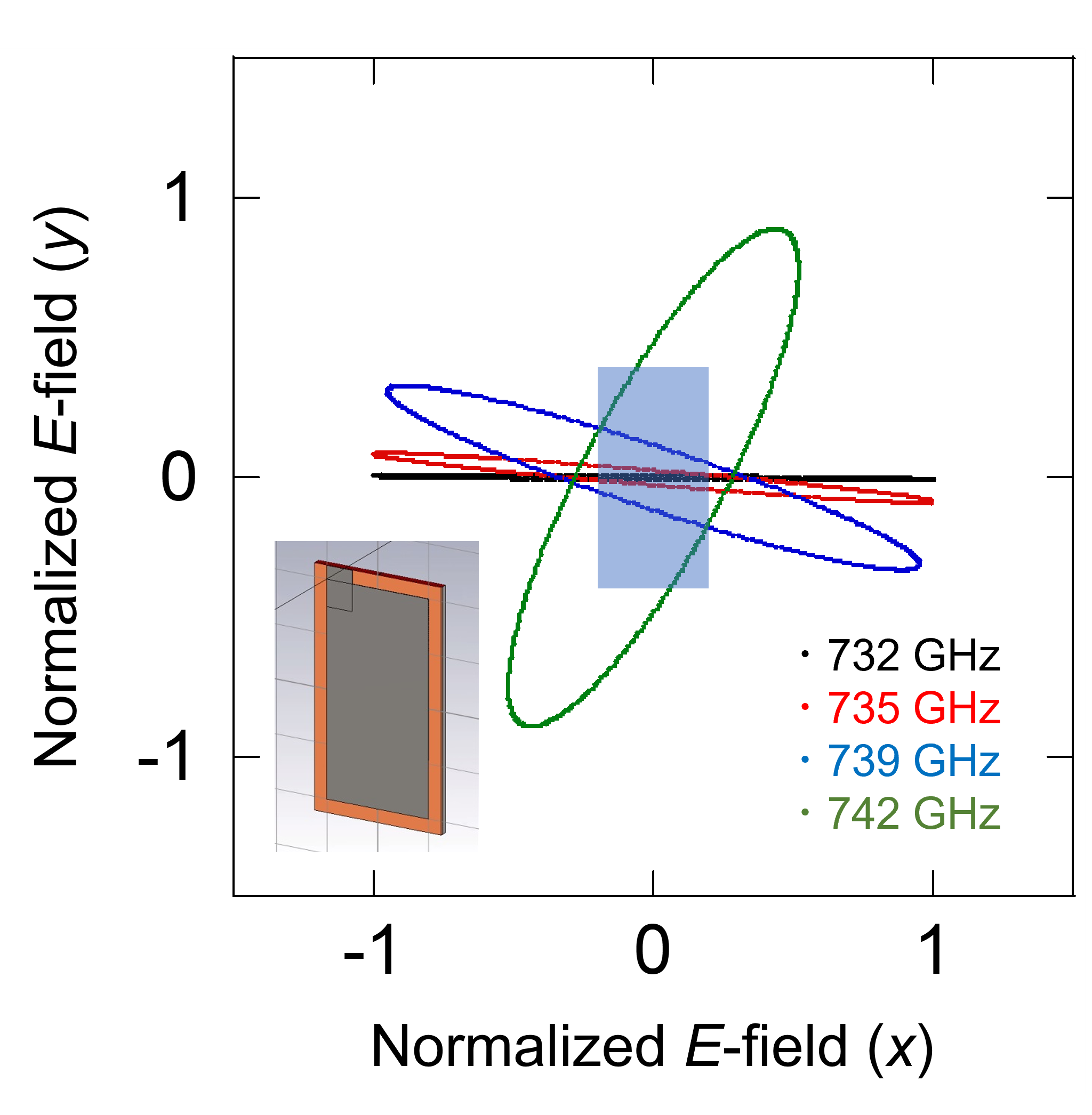}
		\caption{Simulated polarization ellipses in vicinity of antenna resonance frequency.}
	\end{figure}
	
%********************************************************************************************************************************************
%*************************************************************Conclusion**********************************************************************
%********************************************************************************************************************************************

	In conclusion, we studied the polarization characteristics of the THz wave emitted from Bi-2212 whisker crystals and deduced the refractive index using emission mode estimation.
	The refractive index of the whisker crystals was found to be greater than that of bulk single crystals, agreeing with experimental results including our previous report.
	We conducted a 3D electromagnetic simulation to examine the validity of the polarization analysis. 
	The simulation study found that the tilted polarization ellipse appears when the emission point is offset from the cavity resonance point,
	and suggested that the THz emission originates from complex plasma excitation.
	Therefore, the refractive index derived in this study should be treated as a more {\it effective} material parameter to describe the THz emission.
	We believe that this work provides another perspective on the materials and principles for superconducting THz emitters and will contribute to more powerful emissions in the future.

	See the supplementary material for details on sample 1-3, configuration of the simulation, and enlarged image of whisker crystals. 

%****************************************************************************************************************************************************
%**************************************************************Acknowledgement***********************************************************************
%****************************************************************************************************************************************************

	The authors thank Dr. M. Tsujimoto for the invaluable discussions, and S. Harada and T. Ishiyama for all the help.
	This work was supported by JSPS KAKENHI Grant Numbers 20H02606, 20H05644, 19H02177, JSPS Bilateral Program JPJSBP120214602,
	and JST-Mirai Program Grant Number JPMJMI17A2.

\section*{AUTHOR DECLARATIONS}

\section*{Conflict of Interest}
	The authors have no conflicts to disclose.

\section*{Author Contributions}
	{\bf Yoshito Saito}: Investigation (lead); Formal Analysis (lead); Writing - original draft (lead); Writing - review and editing (equal).
	{\bf Itsuhiro Kakeya}: Formal Analysis (supporting); Resources (supporting); Writing - review and editing (equal).
	{\bf Yoshihiko Takano}: Conceptualization (lead); Supervision (lead); Resources (lead); Writing - review and editing (equal).

\section*{DATA AVAILABILITY}
	The data that support the findings of this study are available from the corresponding author upon reasonable request.

\section*{References}

\bibliography{reflist.bib}

\begin{thebibliography}{33}%
\makeatletter
\providecommand \@ifxundefined [1]{%
 \@ifx{#1\undefined}
}%
\providecommand \@ifnum [1]{%
 \ifnum #1\expandafter \@firstoftwo
 \else \expandafter \@secondoftwo
 \fi
}%
\providecommand \@ifx [1]{%
 \ifx #1\expandafter \@firstoftwo
 \else \expandafter \@secondoftwo
 \fi
}%
\providecommand \natexlab [1]{#1}%
\providecommand \enquote  [1]{``#1''}%
\providecommand \bibnamefont  [1]{#1}%
\providecommand \bibfnamefont [1]{#1}%
\providecommand \citenamefont [1]{#1}%
\providecommand \href@noop [0]{\@secondoftwo}%
\providecommand \href [0]{\begingroup \@sanitize@url \@href}%
\providecommand \@href[1]{\@@startlink{#1}\@@href}%
\providecommand \@@href[1]{\endgroup#1\@@endlink}%
\providecommand \@sanitize@url [0]{\catcode `\\12\catcode `\$12\catcode
  `\&12\catcode `\#12\catcode `\^12\catcode `\_12\catcode `\%12\relax}%
\providecommand \@@startlink[1]{}%
\providecommand \@@endlink[0]{}%
\providecommand \url  [0]{\begingroup\@sanitize@url \@url }%
\providecommand \@url [1]{\endgroup\@href {#1}{\urlprefix }}%
\providecommand \urlprefix  [0]{URL }%
\providecommand \Eprint [0]{\href }%
\providecommand \doibase [0]{http://dx.doi.org/}%
\providecommand \selectlanguage [0]{\@gobble}%
\providecommand \bibinfo  [0]{\@secondoftwo}%
\providecommand \bibfield  [0]{\@secondoftwo}%
\providecommand \translation [1]{[#1]}%
\providecommand \BibitemOpen [0]{}%
\providecommand \bibitemStop [0]{}%
\providecommand \bibitemNoStop [0]{.\EOS\space}%
\providecommand \EOS [0]{\spacefactor3000\relax}%
\providecommand \BibitemShut  [1]{\csname bibitem#1\endcsname}%
\let\auto@bib@innerbib\@empty
%</preamble>
\bibitem [{\citenamefont {Kleiner}\ and\ \citenamefont
  {M{\"u}ller}(1994)}]{Kleiner1994}%
  \BibitemOpen
  \bibfield  {author} {\bibinfo {author} {\bibfnamefont {R.}~\bibnamefont
  {Kleiner}}\ and\ \bibinfo {author} {\bibfnamefont {P.}~\bibnamefont
  {M{\"u}ller}},\ }\bibfield  {title} {\enquote {\bibinfo {title} {Intrinsic
  {J}osephson effects in high-${T}_{\rm c}$ superconductors},}\ }\href@noop {}
  {\bibfield  {journal} {\bibinfo  {journal} {Physical Review B}\ }\textbf
  {\bibinfo {volume} {49}},\ \bibinfo {pages} {1327} (\bibinfo {year}
  {1994})}\BibitemShut {NoStop}%
\bibitem [{\citenamefont {Ozyuzer}\ \emph {et~al.}(2007)\citenamefont
  {Ozyuzer}, \citenamefont {Koshelev}, \citenamefont {Kurter}, \citenamefont
  {Gopalsami}, \citenamefont {Li}, \citenamefont {Tachiki}, \citenamefont
  {Kadowaki}, \citenamefont {Yamamoto}, \citenamefont {Minami}, \citenamefont
  {Yamaguchi}, \citenamefont {Tachiki}, \citenamefont {Gray}, \citenamefont
  {Kwok},\ and\ \citenamefont {Welp}}]{Ozyuzer2007}%
  \BibitemOpen
  \bibfield  {author} {\bibinfo {author} {\bibfnamefont {L.}~\bibnamefont
  {Ozyuzer}}, \bibinfo {author} {\bibfnamefont {A.~E.}\ \bibnamefont
  {Koshelev}}, \bibinfo {author} {\bibfnamefont {C.}~\bibnamefont {Kurter}},
  \bibinfo {author} {\bibfnamefont {N.}~\bibnamefont {Gopalsami}}, \bibinfo
  {author} {\bibfnamefont {Q.}~\bibnamefont {Li}}, \bibinfo {author}
  {\bibfnamefont {M.}~\bibnamefont {Tachiki}}, \bibinfo {author} {\bibfnamefont
  {K.}~\bibnamefont {Kadowaki}}, \bibinfo {author} {\bibfnamefont
  {T.}~\bibnamefont {Yamamoto}}, \bibinfo {author} {\bibfnamefont
  {H.}~\bibnamefont {Minami}}, \bibinfo {author} {\bibfnamefont
  {H.}~\bibnamefont {Yamaguchi}}, \bibinfo {author} {\bibfnamefont
  {T.}~\bibnamefont {Tachiki}}, \bibinfo {author} {\bibfnamefont {K.~E.}\
  \bibnamefont {Gray}}, \bibinfo {author} {\bibfnamefont {W.-K.}\ \bibnamefont
  {Kwok}}, \ and\ \bibinfo {author} {\bibfnamefont {U.}~\bibnamefont {Welp}},\
  }\bibfield  {title} {\enquote {\bibinfo {title} {Emission of coherent {TH}z
  radiation from superconductors},}\ }\href {\doibase 10.1126/science.1149802}
  {\bibfield  {journal} {\bibinfo  {journal} {Science}\ }\textbf {\bibinfo
  {volume} {318}},\ \bibinfo {pages} {1291--1293} (\bibinfo {year}
  {2007})}\BibitemShut {NoStop}%
\bibitem [{\citenamefont {Kakeya}\ and\ \citenamefont
  {Wang}(2016)}]{Kakeya2016}%
  \BibitemOpen
  \bibfield  {author} {\bibinfo {author} {\bibfnamefont {I.}~\bibnamefont
  {Kakeya}}\ and\ \bibinfo {author} {\bibfnamefont {H.}~\bibnamefont {Wang}},\
  }\bibfield  {title} {\enquote {\bibinfo {title} {Terahertz-wave emission from
  {B}i2212 intrinsic josephson junctions: {A} review on recent progress},}\
  }\href@noop {} {\bibfield  {journal} {\bibinfo  {journal} {Superconductor
  Science and Technology}\ }\textbf {\bibinfo {volume} {29}},\ \bibinfo {pages}
  {073001} (\bibinfo {year} {2016})}\BibitemShut {NoStop}%
\bibitem [{\citenamefont {Kashiwagi}\ \emph {et~al.}(2017)\citenamefont
  {Kashiwagi}, \citenamefont {Kubo}, \citenamefont {Sakamoto}, \citenamefont
  {Yuasa}, \citenamefont {Tanabe}, \citenamefont {Watanabe}, \citenamefont
  {Tanaka}, \citenamefont {Komori}, \citenamefont {Ota}, \citenamefont {Kuwano}
  \emph {et~al.}}]{Kashiwagi2017}%
  \BibitemOpen
  \bibfield  {author} {\bibinfo {author} {\bibfnamefont {T.}~\bibnamefont
  {Kashiwagi}}, \bibinfo {author} {\bibfnamefont {H.}~\bibnamefont {Kubo}},
  \bibinfo {author} {\bibfnamefont {K.}~\bibnamefont {Sakamoto}}, \bibinfo
  {author} {\bibfnamefont {T.}~\bibnamefont {Yuasa}}, \bibinfo {author}
  {\bibfnamefont {Y.}~\bibnamefont {Tanabe}}, \bibinfo {author} {\bibfnamefont
  {C.}~\bibnamefont {Watanabe}}, \bibinfo {author} {\bibfnamefont
  {T.}~\bibnamefont {Tanaka}}, \bibinfo {author} {\bibfnamefont
  {Y.}~\bibnamefont {Komori}}, \bibinfo {author} {\bibfnamefont
  {R.}~\bibnamefont {Ota}}, \bibinfo {author} {\bibfnamefont {G.}~\bibnamefont
  {Kuwano}},  \emph {et~al.},\ }\bibfield  {title} {\enquote {\bibinfo {title}
  {The present status of high-{T}c superconducting terahertz emitters},}\
  }\href@noop {} {\bibfield  {journal} {\bibinfo  {journal} {Superconductor
  Science and Technology}\ }\textbf {\bibinfo {volume} {30}},\ \bibinfo {pages}
  {074008} (\bibinfo {year} {2017})}\BibitemShut {NoStop}%
\bibitem [{\citenamefont {Kleiner}\ and\ \citenamefont
  {Wang}(2019)}]{Kleiner2019}%
  \BibitemOpen
  \bibfield  {author} {\bibinfo {author} {\bibfnamefont {R.}~\bibnamefont
  {Kleiner}}\ and\ \bibinfo {author} {\bibfnamefont {H.}~\bibnamefont {Wang}},\
  }\bibfield  {title} {\enquote {\bibinfo {title} {Terahertz emission from
  {B}i$_{2}${S}r$_{2}${C}a{C}u$_{2}${O}$_{8+\delta}$ intrinsic {J}osephson
  junction stacks},}\ }\href@noop {} {\bibfield  {journal} {\bibinfo  {journal}
  {Journal of Applied Physics}\ }\textbf {\bibinfo {volume} {126}},\ \bibinfo
  {pages} {171101} (\bibinfo {year} {2019})}\BibitemShut {NoStop}%
\bibitem [{\citenamefont {Saito}\ \emph {et~al.}(2021)\citenamefont {Saito},
  \citenamefont {Adachi}, \citenamefont {Matsumoto}, \citenamefont {Nagao},
  \citenamefont {Fujita}, \citenamefont {Hayama}, \citenamefont {Terashima},
  \citenamefont {Takeya}, \citenamefont {Kakeya},\ and\ \citenamefont
  {Takano}}]{Saito2021}%
  \BibitemOpen
  \bibfield  {author} {\bibinfo {author} {\bibfnamefont {Y.}~\bibnamefont
  {Saito}}, \bibinfo {author} {\bibfnamefont {S.}~\bibnamefont {Adachi}},
  \bibinfo {author} {\bibfnamefont {R.}~\bibnamefont {Matsumoto}}, \bibinfo
  {author} {\bibfnamefont {M.}~\bibnamefont {Nagao}}, \bibinfo {author}
  {\bibfnamefont {S.}~\bibnamefont {Fujita}}, \bibinfo {author} {\bibfnamefont
  {K.}~\bibnamefont {Hayama}}, \bibinfo {author} {\bibfnamefont
  {K.}~\bibnamefont {Terashima}}, \bibinfo {author} {\bibfnamefont
  {H.}~\bibnamefont {Takeya}}, \bibinfo {author} {\bibfnamefont
  {I.}~\bibnamefont {Kakeya}}, \ and\ \bibinfo {author} {\bibfnamefont
  {Y.}~\bibnamefont {Takano}},\ }\bibfield  {title} {\enquote {\bibinfo {title}
  {{TH}z emission from a {B}i$_{2}${S}r$_{2}${C}a{C}u$_{2}${O}$_{8+\delta}$
  cross-whisker junction},}\ }\href@noop {} {\bibfield  {journal} {\bibinfo
  {journal} {Applied Physics Express}\ }\textbf {\bibinfo {volume} {14}},\
  \bibinfo {pages} {033003} (\bibinfo {year} {2021})}\BibitemShut {NoStop}%
\bibitem [{\citenamefont {Cattaneo}\ \emph {et~al.}(2021)\citenamefont
  {Cattaneo}, \citenamefont {Borodianskyi}, \citenamefont {Kalenyuk},\ and\
  \citenamefont {Krasnov}}]{Cattaneo2021}%
  \BibitemOpen
  \bibfield  {author} {\bibinfo {author} {\bibfnamefont {R.}~\bibnamefont
  {Cattaneo}}, \bibinfo {author} {\bibfnamefont {E.~A.}\ \bibnamefont
  {Borodianskyi}}, \bibinfo {author} {\bibfnamefont {A.~A.}\ \bibnamefont
  {Kalenyuk}}, \ and\ \bibinfo {author} {\bibfnamefont {V.~M.}\ \bibnamefont
  {Krasnov}},\ }\bibfield  {title} {\enquote {\bibinfo {title} {Superconducting
  terahertz sources with 12\% power efficiency},}\ }\href@noop {} {\bibfield
  {journal} {\bibinfo  {journal} {Physical Review Applied}\ }\textbf {\bibinfo
  {volume} {16}},\ \bibinfo {pages} {L061001} (\bibinfo {year}
  {2021})}\BibitemShut {NoStop}%
\bibitem [{\citenamefont {Kubo}\ \emph {et~al.}(2010)\citenamefont {Kubo},
  \citenamefont {Takahide}, \citenamefont {Ueda}, \citenamefont {Takano},\ and\
  \citenamefont {Ootuka}}]{Kubo2010}%
  \BibitemOpen
  \bibfield  {author} {\bibinfo {author} {\bibfnamefont {Y.}~\bibnamefont
  {Kubo}}, \bibinfo {author} {\bibfnamefont {Y.}~\bibnamefont {Takahide}},
  \bibinfo {author} {\bibfnamefont {S.}~\bibnamefont {Ueda}}, \bibinfo {author}
  {\bibfnamefont {Y.}~\bibnamefont {Takano}}, \ and\ \bibinfo {author}
  {\bibfnamefont {Y.}~\bibnamefont {Ootuka}},\ }\bibfield  {title} {\enquote
  {\bibinfo {title} {Macroscopic quantum tunneling in a
  {B}i$_{2}${S}r$_{2}${C}a{C}u$_{2}${O}$_{8+\delta}$ single crystalline
  whisker},}\ }\href@noop {} {\bibfield  {journal} {\bibinfo  {journal}
  {Applied physics express}\ }\textbf {\bibinfo {volume} {3}},\ \bibinfo
  {pages} {063104} (\bibinfo {year} {2010})}\BibitemShut {NoStop}%
\bibitem [{\citenamefont {Nagao}\ \emph {et~al.}(2001)\citenamefont {Nagao},
  \citenamefont {Sato}, \citenamefont {Maeda}, \citenamefont {Kim},\ and\
  \citenamefont {Yamashita}}]{Nagao2001}%
  \BibitemOpen
  \bibfield  {author} {\bibinfo {author} {\bibfnamefont {M.}~\bibnamefont
  {Nagao}}, \bibinfo {author} {\bibfnamefont {M.}~\bibnamefont {Sato}},
  \bibinfo {author} {\bibfnamefont {H.}~\bibnamefont {Maeda}}, \bibinfo
  {author} {\bibfnamefont {S.-J.}\ \bibnamefont {Kim}}, \ and\ \bibinfo
  {author} {\bibfnamefont {T.}~\bibnamefont {Yamashita}},\ }\bibfield  {title}
  {\enquote {\bibinfo {title} {Growth and superconducting properties of
  {B}i$_{2}${S}r$_{2}${C}a{C}u$_{2}${O}$_{8+\delta}$ single-crystal whiskers
  using tellurium-doped precursors},}\ }\href@noop {} {\bibfield  {journal}
  {\bibinfo  {journal} {Applied Physics Letters}\ }\textbf {\bibinfo {volume}
  {79}},\ \bibinfo {pages} {2612--2614} (\bibinfo {year} {2001})}\BibitemShut
  {NoStop}%
\bibitem [{\citenamefont {Kishida}\ \emph {et~al.}(2001)\citenamefont
  {Kishida}, \citenamefont {Hirao}, \citenamefont {Kim},\ and\ \citenamefont
  {Yamashita}}]{Kishida2001}%
  \BibitemOpen
  \bibfield  {author} {\bibinfo {author} {\bibfnamefont {S.}~\bibnamefont
  {Kishida}}, \bibinfo {author} {\bibfnamefont {T.}~\bibnamefont {Hirao}},
  \bibinfo {author} {\bibfnamefont {S.~J.}\ \bibnamefont {Kim}}, \ and\
  \bibinfo {author} {\bibfnamefont {T.}~\bibnamefont {Yamashita}},\ }\bibfield
  {title} {\enquote {\bibinfo {title} {Growth of
  {B}i$_{2}${S}r$_{2}${C}a$_{n-1}${C}u$_{n}${O}$_{y}$ superconducting
  whiskers},}\ }\href@noop {} {\bibfield  {journal} {\bibinfo  {journal}
  {Physica C: Superconductivity}\ }\textbf {\bibinfo {volume} {362}},\ \bibinfo
  {pages} {195--199} (\bibinfo {year} {2001})}\BibitemShut {NoStop}%
\bibitem [{\citenamefont {Kadowaki}\ \emph {et~al.}(2008)\citenamefont
  {Kadowaki}, \citenamefont {Yamaguchi}, \citenamefont {Kawamata},
  \citenamefont {Yamamoto}, \citenamefont {Minami}, \citenamefont {Kakeya},
  \citenamefont {Welp}, \citenamefont {Ozyuzer}, \citenamefont {Koshelev},\
  and\ \citenamefont {Kurter}}]{Kadowaki2008}%
  \BibitemOpen
  \bibfield  {author} {\bibinfo {author} {\bibfnamefont {K.}~\bibnamefont
  {Kadowaki}}, \bibinfo {author} {\bibfnamefont {H.}~\bibnamefont {Yamaguchi}},
  \bibinfo {author} {\bibfnamefont {K.}~\bibnamefont {Kawamata}}, \bibinfo
  {author} {\bibfnamefont {T.}~\bibnamefont {Yamamoto}}, \bibinfo {author}
  {\bibfnamefont {H.}~\bibnamefont {Minami}}, \bibinfo {author} {\bibfnamefont
  {I.}~\bibnamefont {Kakeya}}, \bibinfo {author} {\bibfnamefont
  {U.}~\bibnamefont {Welp}}, \bibinfo {author} {\bibfnamefont {L.}~\bibnamefont
  {Ozyuzer}}, \bibinfo {author} {\bibfnamefont {A.}~\bibnamefont {Koshelev}}, \
  and\ \bibinfo {author} {\bibfnamefont {C.}~\bibnamefont {Kurter}},\
  }\bibfield  {title} {\enquote {\bibinfo {title} {Direct observation of
  tetrahertz electromagnetic waves emitted from intrinsic {J}osephson junctions
  in single crystalline {B}i$_{2}${S}r$_{2}${C}a{C}u$_{2}${O}$_{8+\delta}$},}\
  }\href@noop {} {\bibfield  {journal} {\bibinfo  {journal} {Physica C:
  Superconductivity and its applications}\ }\textbf {\bibinfo {volume} {468}},\
  \bibinfo {pages} {634--639} (\bibinfo {year} {2008})}\BibitemShut {NoStop}%
\bibitem [{\citenamefont {Klemm}\ \emph {et~al.}(2011)\citenamefont {Klemm},
  \citenamefont {LaBerge}, \citenamefont {Morley}, \citenamefont {Kashiwagi},
  \citenamefont {Tsujimoto},\ and\ \citenamefont {Kadowaki}}]{Klemm2011}%
  \BibitemOpen
  \bibfield  {author} {\bibinfo {author} {\bibfnamefont {R.~A.}\ \bibnamefont
  {Klemm}}, \bibinfo {author} {\bibfnamefont {E.~R.}\ \bibnamefont {LaBerge}},
  \bibinfo {author} {\bibfnamefont {D.~R.}\ \bibnamefont {Morley}}, \bibinfo
  {author} {\bibfnamefont {T.}~\bibnamefont {Kashiwagi}}, \bibinfo {author}
  {\bibfnamefont {M.}~\bibnamefont {Tsujimoto}}, \ and\ \bibinfo {author}
  {\bibfnamefont {K.}~\bibnamefont {Kadowaki}},\ }\bibfield  {title} {\enquote
  {\bibinfo {title} {Cavity mode waves during terahertz radiation from
  rectangular {B}i$_{2}${S}r$_{2}$({C}a,{Y}){C}u$_{2}${O}$_{8+\delta}$
  mesas},}\ }\href@noop {} {\bibfield  {journal} {\bibinfo  {journal} {Journal
  of Physics. Condensed Matter}\ }\textbf {\bibinfo {volume} {23}} (\bibinfo
  {year} {2011})}\BibitemShut {NoStop}%
\bibitem [{\citenamefont {Benseman}\ \emph {et~al.}(2019)\citenamefont
  {Benseman}, \citenamefont {Koshelev}, \citenamefont {Vlasko-Vlasov},
  \citenamefont {Hao}, \citenamefont {Welp}, \citenamefont {Kwok},
  \citenamefont {Gross}, \citenamefont {Lange}, \citenamefont {Koelle},
  \citenamefont {Kleiner} \emph {et~al.}}]{Benseman2019}%
  \BibitemOpen
  \bibfield  {author} {\bibinfo {author} {\bibfnamefont {T.}~\bibnamefont
  {Benseman}}, \bibinfo {author} {\bibfnamefont {A.}~\bibnamefont {Koshelev}},
  \bibinfo {author} {\bibfnamefont {V.}~\bibnamefont {Vlasko-Vlasov}}, \bibinfo
  {author} {\bibfnamefont {Y.}~\bibnamefont {Hao}}, \bibinfo {author}
  {\bibfnamefont {U.}~\bibnamefont {Welp}}, \bibinfo {author} {\bibfnamefont
  {W.-K.}\ \bibnamefont {Kwok}}, \bibinfo {author} {\bibfnamefont
  {B.}~\bibnamefont {Gross}}, \bibinfo {author} {\bibfnamefont
  {M.}~\bibnamefont {Lange}}, \bibinfo {author} {\bibfnamefont
  {D.}~\bibnamefont {Koelle}}, \bibinfo {author} {\bibfnamefont
  {R.}~\bibnamefont {Kleiner}},  \emph {et~al.},\ }\bibfield  {title} {\enquote
  {\bibinfo {title} {Observation of a two-mode resonant state in a
  {B}i$_{2}${S}r$_{2}${C}a{C}u$_{2}${O}$_{8+\delta}$ mesa device for terahertz
  emission},}\ }\href@noop {} {\bibfield  {journal} {\bibinfo  {journal}
  {Physical Review B}\ }\textbf {\bibinfo {volume} {100}},\ \bibinfo {pages}
  {144503} (\bibinfo {year} {2019})}\BibitemShut {NoStop}%
\bibitem [{\citenamefont {Kobayashi}, \citenamefont {Hayama},\ and\
  \citenamefont {Kakeya}(2022)}]{Kobayashi2022(2)}%
  \BibitemOpen
  \bibfield  {author} {\bibinfo {author} {\bibfnamefont {R.}~\bibnamefont
  {Kobayashi}}, \bibinfo {author} {\bibfnamefont {K.}~\bibnamefont {Hayama}}, \
  and\ \bibinfo {author} {\bibfnamefont {I.}~\bibnamefont {Kakeya}},\
  }\bibfield  {title} {\enquote {\bibinfo {title} {Circuit models of
  simultaneously biased intrinsic {J}osephson junction stacks for terahertz
  radiations in high-bias regime},}\ }\href@noop {} {\bibfield  {journal}
  {\bibinfo  {journal} {Applied Physics Express}\ }\textbf {\bibinfo {volume}
  {15}},\ \bibinfo {pages} {093002} (\bibinfo {year} {2022})}\BibitemShut
  {NoStop}%
\bibitem [{\citenamefont {Saito}\ \emph {et~al.}(2022)\citenamefont {Saito},
  \citenamefont {Maruyama}, \citenamefont {Oda}, \citenamefont {Nagao},
  \citenamefont {Adachi}, \citenamefont {Terashima}, \citenamefont {Tanaka},\
  and\ \citenamefont {Takano}}]{Saito2022}%
  \BibitemOpen
  \bibfield  {author} {\bibinfo {author} {\bibfnamefont {Y.}~\bibnamefont
  {Saito}}, \bibinfo {author} {\bibfnamefont {K.}~\bibnamefont {Maruyama}},
  \bibinfo {author} {\bibfnamefont {K.}~\bibnamefont {Oda}}, \bibinfo {author}
  {\bibfnamefont {M.}~\bibnamefont {Nagao}}, \bibinfo {author} {\bibfnamefont
  {S.}~\bibnamefont {Adachi}}, \bibinfo {author} {\bibfnamefont
  {K.}~\bibnamefont {Terashima}}, \bibinfo {author} {\bibfnamefont
  {I.}~\bibnamefont {Tanaka}}, \ and\ \bibinfo {author} {\bibfnamefont
  {Y.}~\bibnamefont {Takano}},\ }\bibfield  {title} {\enquote {\bibinfo {title}
  {Growth and characterization of
  {B}i$_{2}${S}r$_{2}${C}a$_{1-x}${Y}$_{x}${C}u$_{2}${O}$_{8+\delta}$
  single-crystal whiskers},}\ }\href@noop {} {\bibfield  {journal} {\bibinfo
  {journal} {Japanese Journal of Applied Physics}\ }\textbf {\bibinfo {volume}
  {61}},\ \bibinfo {pages} {063001} (\bibinfo {year} {2022})}\BibitemShut
  {NoStop}%
\bibitem [{\citenamefont {Takano}\ \emph {et~al.}(2001)\citenamefont {Takano},
  \citenamefont {Hatano}, \citenamefont {Fukuyo}, \citenamefont {Ishii},
  \citenamefont {Arisawa}, \citenamefont {Tachiki},\ and\ \citenamefont
  {Togano}}]{Takano2001}%
  \BibitemOpen
  \bibfield  {author} {\bibinfo {author} {\bibfnamefont {Y.}~\bibnamefont
  {Takano}}, \bibinfo {author} {\bibfnamefont {T.}~\bibnamefont {Hatano}},
  \bibinfo {author} {\bibfnamefont {A.}~\bibnamefont {Fukuyo}}, \bibinfo
  {author} {\bibfnamefont {A.}~\bibnamefont {Ishii}}, \bibinfo {author}
  {\bibfnamefont {S.}~\bibnamefont {Arisawa}}, \bibinfo {author} {\bibfnamefont
  {M.}~\bibnamefont {Tachiki}}, \ and\ \bibinfo {author} {\bibfnamefont
  {K.}~\bibnamefont {Togano}},\ }\bibfield  {title} {\enquote {\bibinfo {title}
  {A cross-whiskers junction as a novel fabrication process for intrinsic
  {J}osephson junctions},}\ }\href@noop {} {\bibfield  {journal} {\bibinfo
  {journal} {Superconductor Science and Technology}\ }\textbf {\bibinfo
  {volume} {14}},\ \bibinfo {pages} {765} (\bibinfo {year} {2001})}\BibitemShut
  {NoStop}%
\bibitem [{\citenamefont {Takano}\ \emph {et~al.}(2002)\citenamefont {Takano},
  \citenamefont {Hatano}, \citenamefont {Fukuyo}, \citenamefont {Ishii},
  \citenamefont {Ohmori}, \citenamefont {Arisawa}, \citenamefont {Togano},\
  and\ \citenamefont {Tachiki}}]{Takano2002}%
  \BibitemOpen
  \bibfield  {author} {\bibinfo {author} {\bibfnamefont {Y.}~\bibnamefont
  {Takano}}, \bibinfo {author} {\bibfnamefont {T.}~\bibnamefont {Hatano}},
  \bibinfo {author} {\bibfnamefont {A.}~\bibnamefont {Fukuyo}}, \bibinfo
  {author} {\bibfnamefont {A.}~\bibnamefont {Ishii}}, \bibinfo {author}
  {\bibfnamefont {M.}~\bibnamefont {Ohmori}}, \bibinfo {author} {\bibfnamefont
  {S.}~\bibnamefont {Arisawa}}, \bibinfo {author} {\bibfnamefont
  {K.}~\bibnamefont {Togano}}, \ and\ \bibinfo {author} {\bibfnamefont
  {M.}~\bibnamefont {Tachiki}},\ }\bibfield  {title} {\enquote {\bibinfo
  {title} {d-like symmetry of the order parameter and intrinsic {J}osephson
  effects in {B}i$_{2}${S}r$_{2}$({C}a,{Y}){C}u$_{2}${O}$_{8+\delta}$
  cross-whisker junctions},}\ }\href@noop {} {\bibfield  {journal} {\bibinfo
  {journal} {Physical Review B}\ }\textbf {\bibinfo {volume} {65}},\ \bibinfo
  {pages} {140513} (\bibinfo {year} {2002})}\BibitemShut {NoStop}%
\bibitem [{\citenamefont {Yurgens}(2011)}]{Yurgens2011}%
  \BibitemOpen
  \bibfield  {author} {\bibinfo {author} {\bibfnamefont {A.}~\bibnamefont
  {Yurgens}},\ }\bibfield  {title} {\enquote {\bibinfo {title} {Temperature
  distribution in a large {B}i$_{2}${S}r$_{2}${C}a{C}u$_{2}${O}$_{8+\delta}$
  mesa},}\ }\href@noop {} {\bibfield  {journal} {\bibinfo  {journal} {Physical
  Review B}\ }\textbf {\bibinfo {volume} {83}},\ \bibinfo {pages} {184501}
  (\bibinfo {year} {2011})}\BibitemShut {NoStop}%
\bibitem [{\citenamefont {Demirhan}\ \emph {et~al.}(2015)\citenamefont
  {Demirhan}, \citenamefont {Saglam}, \citenamefont {Turkoglu}, \citenamefont
  {Alaboz}, \citenamefont {Ozyuzer}, \citenamefont {Miyakawa},\ and\
  \citenamefont {Kadowaki}}]{Demirhan2015}%
  \BibitemOpen
  \bibfield  {author} {\bibinfo {author} {\bibfnamefont {Y.}~\bibnamefont
  {Demirhan}}, \bibinfo {author} {\bibfnamefont {H.}~\bibnamefont {Saglam}},
  \bibinfo {author} {\bibfnamefont {F.}~\bibnamefont {Turkoglu}}, \bibinfo
  {author} {\bibfnamefont {H.}~\bibnamefont {Alaboz}}, \bibinfo {author}
  {\bibfnamefont {L.}~\bibnamefont {Ozyuzer}}, \bibinfo {author} {\bibfnamefont
  {N.}~\bibnamefont {Miyakawa}}, \ and\ \bibinfo {author} {\bibfnamefont
  {K.}~\bibnamefont {Kadowaki}},\ }\bibfield  {title} {\enquote {\bibinfo
  {title} {Area dependence and influence of crystal inhomogeneity on
  superconducting properties of bi2212 mesa structures},}\ }\href@noop {}
  {\bibfield  {journal} {\bibinfo  {journal} {Vacuum}\ }\textbf {\bibinfo
  {volume} {120}},\ \bibinfo {pages} {89--94} (\bibinfo {year}
  {2015})}\BibitemShut {NoStop}%
\bibitem [{\citenamefont {Slack}(1962)}]{Slack1962}%
  \BibitemOpen
  \bibfield  {author} {\bibinfo {author} {\bibfnamefont {G.~A.}\ \bibnamefont
  {Slack}},\ }\bibfield  {title} {\enquote {\bibinfo {title} {Thermal
  conductivity of {M}g{O}, {A}l$_{2}${O}$_{3}$, {M}g{A}l$_{2}${O}$_{4}$, and
  {F}e$_{3}${O}$_{4}$ crystals from 3$^{\circ}$to 300$^{\circ}$ k},}\
  }\href@noop {} {\bibfield  {journal} {\bibinfo  {journal} {Physical Review}\
  }\textbf {\bibinfo {volume} {126}},\ \bibinfo {pages} {427} (\bibinfo {year}
  {1962})}\BibitemShut {NoStop}%
\bibitem [{\citenamefont {Fujishiro}\ \emph {et~al.}(1994)\citenamefont
  {Fujishiro}, \citenamefont {Ikebe}, \citenamefont {Naito}, \citenamefont
  {Matsukawa}, \citenamefont {Noto}, \citenamefont {Shigaki}, \citenamefont
  {Shibutani}, \citenamefont {Hayashi},\ and\ \citenamefont
  {Ogawa}}]{Fujishiro1994}%
  \BibitemOpen
  \bibfield  {author} {\bibinfo {author} {\bibfnamefont {H.}~\bibnamefont
  {Fujishiro}}, \bibinfo {author} {\bibfnamefont {M.}~\bibnamefont {Ikebe}},
  \bibinfo {author} {\bibfnamefont {T.}~\bibnamefont {Naito}}, \bibinfo
  {author} {\bibfnamefont {M.}~\bibnamefont {Matsukawa}}, \bibinfo {author}
  {\bibfnamefont {K.}~\bibnamefont {Noto}}, \bibinfo {author} {\bibfnamefont
  {I.}~\bibnamefont {Shigaki}}, \bibinfo {author} {\bibfnamefont
  {K.}~\bibnamefont {Shibutani}}, \bibinfo {author} {\bibfnamefont
  {S.}~\bibnamefont {Hayashi}}, \ and\ \bibinfo {author} {\bibfnamefont
  {R.}~\bibnamefont {Ogawa}},\ }\bibfield  {title} {\enquote {\bibinfo {title}
  {Thermal conductivity of {B}i-2212 single crystals prepared by {TSFZ}
  method},}\ }\href@noop {} {\bibfield  {journal} {\bibinfo  {journal} {Physica
  C: Superconductivity}\ }\textbf {\bibinfo {volume} {235}},\ \bibinfo {pages}
  {1533--1534} (\bibinfo {year} {1994})}\BibitemShut {NoStop}%
\bibitem [{\citenamefont {Collet}(2005)}]{Collet2005}%
  \BibitemOpen
  \bibfield  {author} {\bibinfo {author} {\bibfnamefont {E.}~\bibnamefont
  {Collet}},\ }\bibfield  {title} {\enquote {\bibinfo {title} {Field guide to
  polarization, spie vol},}\ }\href@noop {} {\bibfield  {journal} {\bibinfo
  {journal} {FG05}\ } (\bibinfo {year} {2005})}\BibitemShut {NoStop}%
\bibitem [{\citenamefont {Revcolevschi}\ and\ \citenamefont
  {Deutscher}(1996)}]{Revcolevschi1996}%
  \BibitemOpen
  \bibfield  {author} {\bibinfo {author} {\bibfnamefont {A.}~\bibnamefont
  {Revcolevschi}}\ and\ \bibinfo {author} {\bibfnamefont {G.}~\bibnamefont
  {Deutscher}},\ }\href@noop {} {\emph {\bibinfo {title} {Coherence in high
  temperature superconductors}}}\ (\bibinfo  {publisher} {World Scientific},\
  \bibinfo {year} {1996})\BibitemShut {NoStop}%
\bibitem [{\citenamefont {Kakeya}\ \emph {et~al.}(1998)\citenamefont {Kakeya},
  \citenamefont {Kindo}, \citenamefont {Kadowaki}, \citenamefont {Takahashi},\
  and\ \citenamefont {Mochiku}}]{Kakeya1998}%
  \BibitemOpen
  \bibfield  {author} {\bibinfo {author} {\bibfnamefont {I.}~\bibnamefont
  {Kakeya}}, \bibinfo {author} {\bibfnamefont {K.}~\bibnamefont {Kindo}},
  \bibinfo {author} {\bibfnamefont {K.}~\bibnamefont {Kadowaki}}, \bibinfo
  {author} {\bibfnamefont {S.}~\bibnamefont {Takahashi}}, \ and\ \bibinfo
  {author} {\bibfnamefont {T.}~\bibnamefont {Mochiku}},\ }\bibfield  {title}
  {\enquote {\bibinfo {title} {Mode separation of the josephson plasma in
  {B}i$_{2}${S}r$_{2}${C}a{C}u$_{2}${O}$_{8+\delta}$},}\ }\href@noop {}
  {\bibfield  {journal} {\bibinfo  {journal} {Physical review B}\ }\textbf
  {\bibinfo {volume} {57}},\ \bibinfo {pages} {3108} (\bibinfo {year}
  {1998})}\BibitemShut {NoStop}%
\bibitem [{\citenamefont {Kashiwagi}\ \emph {et~al.}(2011)\citenamefont
  {Kashiwagi}, \citenamefont {Tsujimoto}, \citenamefont {Yamamoto},
  \citenamefont {Minami}, \citenamefont {Yamaki}, \citenamefont {Delfanazari},
  \citenamefont {Deguchi}, \citenamefont {Orita}, \citenamefont {Koike},
  \citenamefont {Nakayama} \emph {et~al.}}]{Kashiwagi2011}%
  \BibitemOpen
  \bibfield  {author} {\bibinfo {author} {\bibfnamefont {T.}~\bibnamefont
  {Kashiwagi}}, \bibinfo {author} {\bibfnamefont {M.}~\bibnamefont
  {Tsujimoto}}, \bibinfo {author} {\bibfnamefont {T.}~\bibnamefont {Yamamoto}},
  \bibinfo {author} {\bibfnamefont {H.}~\bibnamefont {Minami}}, \bibinfo
  {author} {\bibfnamefont {K.}~\bibnamefont {Yamaki}}, \bibinfo {author}
  {\bibfnamefont {K.}~\bibnamefont {Delfanazari}}, \bibinfo {author}
  {\bibfnamefont {K.}~\bibnamefont {Deguchi}}, \bibinfo {author} {\bibfnamefont
  {N.}~\bibnamefont {Orita}}, \bibinfo {author} {\bibfnamefont
  {T.}~\bibnamefont {Koike}}, \bibinfo {author} {\bibfnamefont
  {R.}~\bibnamefont {Nakayama}},  \emph {et~al.},\ }\bibfield  {title}
  {\enquote {\bibinfo {title} {High temperature superconductor terahertz
  emitters: {F}undamental physics and its applications},}\ }\href@noop {}
  {\bibfield  {journal} {\bibinfo  {journal} {Japanese Journal of Applied
  Physics}\ }\textbf {\bibinfo {volume} {51}},\ \bibinfo {pages} {010113}
  (\bibinfo {year} {2011})}\BibitemShut {NoStop}%
\bibitem [{\citenamefont {Mochiku}\ and\ \citenamefont
  {Kadowaki}(1994)}]{Mochiku1994}%
  \BibitemOpen
  \bibfield  {author} {\bibinfo {author} {\bibfnamefont {T.}~\bibnamefont
  {Mochiku}}\ and\ \bibinfo {author} {\bibfnamefont {K.}~\bibnamefont
  {Kadowaki}},\ }\bibfield  {title} {\enquote {\bibinfo {title} {Growth and
  properties of {B}i$_{2}${S}r$_{2}$({C}a,{Y}){C}u$_{2}${O}$_{8+\delta}$ single
  crystals},}\ }\href@noop {} {\bibfield  {journal} {\bibinfo  {journal}
  {Physica C: Superconductivity}\ }\textbf {\bibinfo {volume} {235}},\ \bibinfo
  {pages} {523--524} (\bibinfo {year} {1994})}\BibitemShut {NoStop}%
\bibitem [{\citenamefont {Motohashi}\ \emph {et~al.}(2000)\citenamefont
  {Motohashi}, \citenamefont {Shimoyama}, \citenamefont {Kitazawa},
  \citenamefont {Kishio}, \citenamefont {Kojima}, \citenamefont {Uchida},\ and\
  \citenamefont {Tajima}}]{Motohashi2000}%
  \BibitemOpen
  \bibfield  {author} {\bibinfo {author} {\bibfnamefont {T.}~\bibnamefont
  {Motohashi}}, \bibinfo {author} {\bibfnamefont {J.}~\bibnamefont
  {Shimoyama}}, \bibinfo {author} {\bibfnamefont {K.}~\bibnamefont {Kitazawa}},
  \bibinfo {author} {\bibfnamefont {K.}~\bibnamefont {Kishio}}, \bibinfo
  {author} {\bibfnamefont {K.}~\bibnamefont {Kojima}}, \bibinfo {author}
  {\bibfnamefont {S.}~\bibnamefont {Uchida}}, \ and\ \bibinfo {author}
  {\bibfnamefont {S.}~\bibnamefont {Tajima}},\ }\bibfield  {title} {\enquote
  {\bibinfo {title} {Observation of the {J}osephson plasma reflectivity edge in
  the infrared region in {B}i-based superconducting cuprates},}\ }\href@noop {}
  {\bibfield  {journal} {\bibinfo  {journal} {Physical Review B}\ }\textbf
  {\bibinfo {volume} {61}},\ \bibinfo {pages} {R9269} (\bibinfo {year}
  {2000})}\BibitemShut {NoStop}%
\bibitem [{\citenamefont {Tajima}\ \emph {et~al.}(1993)\citenamefont {Tajima},
  \citenamefont {Gu}, \citenamefont {Miyamoto}, \citenamefont {Odagawa},\ and\
  \citenamefont {Koshizuka}}]{Tajima1993}%
  \BibitemOpen
  \bibfield  {author} {\bibinfo {author} {\bibfnamefont {S.}~\bibnamefont
  {Tajima}}, \bibinfo {author} {\bibfnamefont {G.}~\bibnamefont {Gu}}, \bibinfo
  {author} {\bibfnamefont {S.}~\bibnamefont {Miyamoto}}, \bibinfo {author}
  {\bibfnamefont {A.}~\bibnamefont {Odagawa}}, \ and\ \bibinfo {author}
  {\bibfnamefont {N.}~\bibnamefont {Koshizuka}},\ }\bibfield  {title} {\enquote
  {\bibinfo {title} {Optical evidence for strong anisotropy in the normal and
  superconducting states in {B}i$_{2}${S}r$_{2}${C}a{C}u$_{2}${O}$_{8+z}$},}\
  }\href@noop {} {\bibfield  {journal} {\bibinfo  {journal} {Physical Review
  B}\ }\textbf {\bibinfo {volume} {48}},\ \bibinfo {pages} {16164} (\bibinfo
  {year} {1993})}\BibitemShut {NoStop}%
\bibitem [{\citenamefont {Gaifullin}\ \emph {et~al.}(2000)\citenamefont
  {Gaifullin}, \citenamefont {Matsuda}, \citenamefont {Chikumoto},
  \citenamefont {Shimoyama},\ and\ \citenamefont {Kishio}}]{Gaifullin2000}%
  \BibitemOpen
  \bibfield  {author} {\bibinfo {author} {\bibfnamefont {M.}~\bibnamefont
  {Gaifullin}}, \bibinfo {author} {\bibfnamefont {Y.}~\bibnamefont {Matsuda}},
  \bibinfo {author} {\bibfnamefont {N.}~\bibnamefont {Chikumoto}}, \bibinfo
  {author} {\bibfnamefont {J.}~\bibnamefont {Shimoyama}}, \ and\ \bibinfo
  {author} {\bibfnamefont {K.}~\bibnamefont {Kishio}},\ }\bibfield  {title}
  {\enquote {\bibinfo {title} {Abrupt change of {J}osephson plasma frequency at
  the phase boundary of the {B}ragg glass in
  {B}i$_{2}${S}r$_{2}${C}a{C}u$_{2}${O}$_{8+\delta}$},}\ }\href@noop {}
  {\bibfield  {journal} {\bibinfo  {journal} {Physical Review Letters}\
  }\textbf {\bibinfo {volume} {84}},\ \bibinfo {pages} {2945} (\bibinfo {year}
  {2000})}\BibitemShut {NoStop}%
\bibitem [{\citenamefont {Elarabi}\ \emph {et~al.}(2017)\citenamefont
  {Elarabi}, \citenamefont {Yoshioka}, \citenamefont {Tsujimoto},\ and\
  \citenamefont {Kakeya}}]{Elarabi2017}%
  \BibitemOpen
  \bibfield  {author} {\bibinfo {author} {\bibfnamefont {A.}~\bibnamefont
  {Elarabi}}, \bibinfo {author} {\bibfnamefont {Y.}~\bibnamefont {Yoshioka}},
  \bibinfo {author} {\bibfnamefont {M.}~\bibnamefont {Tsujimoto}}, \ and\
  \bibinfo {author} {\bibfnamefont {I.}~\bibnamefont {Kakeya}},\ }\bibfield
  {title} {\enquote {\bibinfo {title} {Monolithic superconducting emitter of
  tunable circularly polarized terahertz radiation},}\ }\href@noop {}
  {\bibfield  {journal} {\bibinfo  {journal} {Physical Review Applied}\
  }\textbf {\bibinfo {volume} {8}},\ \bibinfo {pages} {064034} (\bibinfo {year}
  {2017})}\BibitemShut {NoStop}%
\bibitem [{\citenamefont {Elarabi}\ \emph {et~al.}(2018)\citenamefont
  {Elarabi}, \citenamefont {Yoshioka}, \citenamefont {Tsujimoto},\ and\
  \citenamefont {Kakeya}}]{Elarabi2018}%
  \BibitemOpen
  \bibfield  {author} {\bibinfo {author} {\bibfnamefont {A.}~\bibnamefont
  {Elarabi}}, \bibinfo {author} {\bibfnamefont {Y.}~\bibnamefont {Yoshioka}},
  \bibinfo {author} {\bibfnamefont {M.}~\bibnamefont {Tsujimoto}}, \ and\
  \bibinfo {author} {\bibfnamefont {I.}~\bibnamefont {Kakeya}},\ }\bibfield
  {title} {\enquote {\bibinfo {title} {Circularly polarized terahertz radiation
  monolithically generated by cylindrical mesas of intrinsic {J}osephson
  junctions},}\ }\href@noop {} {\bibfield  {journal} {\bibinfo  {journal}
  {Applied Physics Letters}\ }\textbf {\bibinfo {volume} {113}},\ \bibinfo
  {pages} {052601} (\bibinfo {year} {2018})}\BibitemShut {NoStop}%
\bibitem [{\citenamefont {Wang}\ \emph {et~al.}(2010)\citenamefont {Wang},
  \citenamefont {Gu{\'e}non}, \citenamefont {Gross}, \citenamefont {Yuan},
  \citenamefont {Jiang}, \citenamefont {Zhong}, \citenamefont {Gr{\"u}nzweig},
  \citenamefont {Iishi}, \citenamefont {Wu}, \citenamefont {Hatano} \emph
  {et~al.}}]{Wang2010}%
  \BibitemOpen
  \bibfield  {author} {\bibinfo {author} {\bibfnamefont {H.}~\bibnamefont
  {Wang}}, \bibinfo {author} {\bibfnamefont {S.}~\bibnamefont {Gu{\'e}non}},
  \bibinfo {author} {\bibfnamefont {B.}~\bibnamefont {Gross}}, \bibinfo
  {author} {\bibfnamefont {J.}~\bibnamefont {Yuan}}, \bibinfo {author}
  {\bibfnamefont {Z.}~\bibnamefont {Jiang}}, \bibinfo {author} {\bibfnamefont
  {Y.}~\bibnamefont {Zhong}}, \bibinfo {author} {\bibfnamefont
  {M.}~\bibnamefont {Gr{\"u}nzweig}}, \bibinfo {author} {\bibfnamefont
  {A.}~\bibnamefont {Iishi}}, \bibinfo {author} {\bibfnamefont
  {P.}~\bibnamefont {Wu}}, \bibinfo {author} {\bibfnamefont {T.}~\bibnamefont
  {Hatano}},  \emph {et~al.},\ }\bibfield  {title} {\enquote {\bibinfo {title}
  {Coherent terahertz emission of intrinsic josephson {J}unction stacks in the
  hot spot regime},}\ }\href@noop {} {\bibfield  {journal} {\bibinfo  {journal}
  {Physical review letters}\ }\textbf {\bibinfo {volume} {105}},\ \bibinfo
  {pages} {057002} (\bibinfo {year} {2010})}\BibitemShut {NoStop}%
\bibitem [{\citenamefont {Tsujimoto}\ \emph {et~al.}(2020)\citenamefont
  {Tsujimoto}, \citenamefont {Fujita}, \citenamefont {Kuwano}, \citenamefont
  {Maeda}, \citenamefont {Elarabi}, \citenamefont {Hawecker}, \citenamefont
  {Tignon}, \citenamefont {Mangeney}, \citenamefont {Dhillon},\ and\
  \citenamefont {Kakeya}}]{Tsujimoto2020}%
  \BibitemOpen
  \bibfield  {author} {\bibinfo {author} {\bibfnamefont {M.}~\bibnamefont
  {Tsujimoto}}, \bibinfo {author} {\bibfnamefont {S.}~\bibnamefont {Fujita}},
  \bibinfo {author} {\bibfnamefont {G.}~\bibnamefont {Kuwano}}, \bibinfo
  {author} {\bibfnamefont {K.}~\bibnamefont {Maeda}}, \bibinfo {author}
  {\bibfnamefont {A.}~\bibnamefont {Elarabi}}, \bibinfo {author} {\bibfnamefont
  {J.}~\bibnamefont {Hawecker}}, \bibinfo {author} {\bibfnamefont
  {J.}~\bibnamefont {Tignon}}, \bibinfo {author} {\bibfnamefont
  {J.}~\bibnamefont {Mangeney}}, \bibinfo {author} {\bibfnamefont
  {S.}~\bibnamefont {Dhillon}}, \ and\ \bibinfo {author} {\bibfnamefont
  {I.}~\bibnamefont {Kakeya}},\ }\bibfield  {title} {\enquote {\bibinfo {title}
  {Mutually synchronized macroscopic {J}osephson oscillations demonstrated by
  polarization analysis of superconducting terahertz emitters},}\ }\href@noop
  {} {\bibfield  {journal} {\bibinfo  {journal} {Physical Review Applied}\
  }\textbf {\bibinfo {volume} {13}},\ \bibinfo {pages} {051001} (\bibinfo
  {year} {2020})}\BibitemShut {NoStop}%
\end{thebibliography}%

\end{document}